# Computer simulations of the extinction of megafauna

Ron W. Nielsen[1]

**Abstract**. Computer simulations carried out by Alroy (2001) are examined. Contrary to the claim of their Author, there is no convincing evidence that the extinction of megafauna was caused by humans. Intentionally or unintentionally, attempt was made to force the human-induced extinction of megafauna by assuming an absurdly fast growth of the hypothetical human population in North America. The assumed growth rate was around two-orders of magnitude larger than normally expected. It is well known that the past growth of human population was slow, but in these simulations, the growth of the hypothetical human population was unreasonably fast. However, even under this unreasonable assumption about the growth of human population, computer simulations do not support the postulate of human-assisted extinction of megafauna because there is no clear correlation between the growth of the hypothetical human population and the distributions describing the decline in the population of megafauna.

## Introduction

Mathematical modelling makes sense only if model calculations are compared with data; otherwise it is just a story, which might or might not be true. It is like translating a story from one language to another. We might tell a story based on our imagination and translate it into mathematical language but just because we have now presented our story in a mathematical language it does not mean that we have endowed it with a scientific authority. It is still just a story. We have to compare our story with data and only then can we tell whether our story (our model) is just a fiction or whether it has scientific value. We should also not be mesmerised by computer outputs. If we feed nonsense into a computer we can expect nonsense in return. Computers are not endowed with a scientific authority.

In my two earlier publications (Nielsen, 2017a, 2017b) I have analysed the study of Barnosky (2008) and his claim that population data support the postulate of human-assisted extinction of megafauna. I have shown that Barnosky was using fabricated "data" for the growth of human population. I have shown that these "data" were constructed by Hern (1999). They were so obviously incorrect that they should have never been used. Even Hern clearly indicated in his publication that his "data" were constructed by him. However, I have also shown that contrary to the claim made by Barnosky, even these fabricated "data" do not support the postulate of the human-assisted extinction of megafauna.

In the document presented here, I am going to discuss model calculations carried out by Alroy (2001). I am going to show that contrary to his claim, *his model does not support the postulate of the human-assisted extinction of megafauna* for two reasons: (1) there is *no correlation* between the model-generated distribution describing the growth of the hypothetical human population and the distributions describing the decline in the population of megafauna and (2) his computer-generated, fast-increasing, growth of the hypothetical human population in North America *does not describe the growth of human population.*

---

[1]AKA Jan Nurzynski, Environmental Futures Research Institute, Griffith University, Qld, 4222, Australia. r.nielsen@griffith.edu.au; ronwnielsen@gmail.com;

# Computer simulations

Alroy (2001) simulated the growth of population and the decline in the population of megafauna in North America starting from 14,000 BP (years before present). He explains: "I simulate human population growth, hunting patterns, and the population dynamics of 41 large mammalian herbivores – 30 of them now extinct – across the Pleistocene-Holocene transition" (Alroy, 2001, p. 1893).

In his Fig. 1 (Alroy, 2001, p. 1895), he shows two types of distributions: the distribution describing the growth of human population and a cluster of distributions describing the population of megafauna. The essential features of his diagram are reproduced here as Figure 1, which presents only one distribution describing the population of megafauna. Nearly all other distributions are virtually identical. They are characterised by a constant size of the population of megafauna for about 500 years followed by a sudden decline in their populations. The size of various species during the time when their population was supposed to have remained constant (for about 500 years) ranges from 1000 to over 1,000,000. The right-hand side of his Fig. 1, the side which is not reproduced in my diagram, shows that the computer-generated distribution describing the growth of the hypothetical human population in North America reaches quickly a constant value, which remains constant until the present time. This is also a questionable feature but I am not going to discuss it here. The distribution describing the growth of the hypothetical human population in North America, shown in Figure 1 and in Alroy's Fig. 1, is his preferred distribution, which he describes as trial 8.

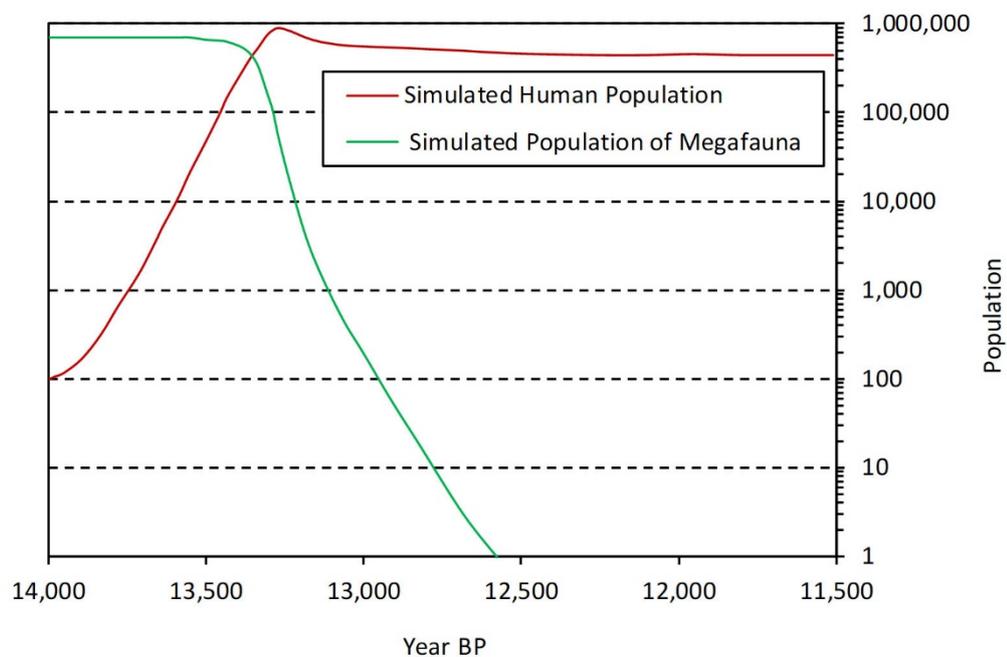

**Figure 1.** *Example of computer simulations of the growth of human population and of the decline in the population of megafauna (Alroy, 2001). There is no clear correlation between the growth of the hypothetical human population and the decline in the population of megafauna. Furthermore, the simulated growth of the hypothetical human population was so absurdly fast that it is scientifically unacceptable. It increased by a factor of around 10,000 in just 700 years. For the world population the same increase was after 1,000,000 years! (Deevey, 1960). However, if we include also estimates of Birdsell (1972) and Hassan (2002), the increase was only by a factor of 2,300 in 1,000,000 years. Alroy's simulations are so far off the mark that they have to be rejected.*

There are certain questions about the overall shapes of these distributions but I am not going to discuss them because there are two other issues that make these computer simulations already scientifically unacceptable, and in particular just one and glaringly obvious issue: the unacceptably fast growth of the hypothetical human population. It should be also noted that Alroy's model calculations *were never compared with data describing the growth of human population*. Thus, the fundamental and essential step in the scientific research has not been taken.

Alroy's simulations do not simulate the human-assisted extinction of megafauna, for two reasons. *First*, there is no correlation between the distributions describing the population of megafauna and the distribution describing the growth of the hypothetical human population. *Second*, the distribution describing the growth of the hypothetical human population increases unacceptably fast and it has an overall unacceptable shape. It is so unlike the distributions describing the growth of human population, published before or at the time of the publication of Alroy's study (Deevey, 1960; Maddison, 2001; McEvedy & Jones, 1978; von Foerster, Mora & Amiot, 1960), that it has to be rejected as irrelevant and incorrect.

Alroy's simulations do not simulate processes that might have taken place in the real world but rather a process in the world of fiction, in the world where the hypothetical human population was increasing excessively fast over a short time to reach quickly a constant value, which remained constant for many thousands of years, a strange world characterised by the unfamiliar interaction of the hypothetical humans with the environment. Alroy might not have had data describing the growth of human population in North America but he had other relevant data (Deevey, 1960; Maddison, 2001; McEvedy & Jones, 1978; von Foerster, Mora & Amiot, 1960), which would have shown to him that his simulations of the growth of the hypothetical human population were clearly incorrect.

There appears to have been a serious fundamental problem with the basic assumptions used in these simulations if the calculations produced such an unacceptable distribution for the growth of the alleged human population. Maybe the simulated growth of the hypothetical human population had to be *forced* to be so absurdly fast in order to generate the extinction of megafauna. Maybe it was a purposeful manipulation of evidence or maybe it was simply the unskilful way of carrying out scientific research. In any case, Alroy does not give any evidence, even if this is just only simulated evidence, that the extinction of megafauna was caused by humans.

In the example shown in Figure 1 human population is assumed to have been increasing rapidly, with the growth rate of 1.66% per annum (Alroy, 2001, p. 1894), unbelievably fast, at a breath-taking speed. At this rate, the size of human population was doubling every 42 years. The size of this hypothetical population would have increased from the original size of 100 to 1,000,000,000 (one billion!) in only *1250* years. Now compare this fast increase with what was happening in the real world. The size of the world population reached one billion after about *1,000,000* years even though the starting value was already around 125,000 (Deevey, 1960). Should it not make Alroy uncomfortable about his far-fetched assumptions?

Alroy's hypothetical human population was growing absurdly fast but he describes it as slow ["Slow human growth rates" (Alroy, 2001, p. 1893)]. It was not slow; it was excessively fast. Whatever he was simulating it was not the growth of human population.

We should notice that even this fast-increasing human population had no impact on the population of megafauna, which remained constant for about 500 years. According to Alroy, the population of some species of megafauna was small during that time. We would expect to see some impact by the fast-growing human population on these small populations of megafauna but the simulations produce not even the slightest impact during the first 500 years. Then suddenly, and for no apparent reason, the populations of megafauna started to decline rapidly, like an avalanche, indiscriminately and independently on whether their populations were counted in around millions or in a few thousands. Human impact was suddenly everywhere. Furthermore, this rapid decline in the populations of megafauna is not reflected in the growth of the hypothetical human population. Distributions describing the decline in the population of megafauna and the distribution describing the growth of population are independent.

It is true that for a good part of the initial 500 years the population of these hypothetical humans was small and they did not have any significant impact on reducing the population of megafauna, but why was there such a massive and *indiscriminate* decline in the population of megafauna towards the end of these first 500 years? How to explain this sudden killing spree and why was it equally efficient for the small and for the large populations of megafauna.? Whether their initial population was around 1000 or around 1,000,000 they followed the same fate at approximately the same time. Their populations were suddenly reduced within approximately the same section of time.

Alroy's claim that "This fully mechanistic model accounts for megafaunal extinction without invoking climate change and secondary ecological effects" (Alroy, 2001, p. 1893) *is not supported even by his calculations*. Not only do they generate a strange and unacceptably fast growth of the alleged human population but also, they do not even demonstrate a clear close correlation between this distribution and the distributions describing the extinction of megafauna.

Computer simulations tell one story while Alroy tells another. The two stories are diametrically different. Computer simulations show no correlation between the extinction of megafauna and the growth of the alleged human population but Alroy claims that his "model accounts for megafaunal extinction without invoking climate change and secondary ecological effects."

The growth of the hypothetical human population was approximately exponential because the distribution during that time is represented by an approximately straight line in this semilogarithmic display, and it was a fast exponential growth because it was characterised by a large growth rate of 1.66%. A quick check of the publication of Deevey (1960) or Kremer (1993), both available to Alroy, would have demonstrated to him that the assumed growth rate of 1.66% was manifestly incorrect and, consequently, unacceptable. This growth rate is nowhere near an acceptable and realistic estimate. The growth rate assumed by Alroy is so incorrect that it renders the whole model unacceptable.

Kremer did not analyse population data but he carried out approximate calculations of the growth rates. He gives the growth rate of 0.0045% in 10,000 BC (or in 11.950 BP if we count the BP years from 1950) and 0.0031% in 25,000 BC (Kremer, 1993, p. 683). Even if we use these approximate values, we can see clearly that Alroy's value of 1.66% is unacceptable. The discrepancy is so large and so obvious that it should have been noticed by Alroy and certainly it should have been noticed by the editing team of *Science* but the absence of the correlation between the alleged growth of human population and the alleged decline in the growth of the population of megafauna should also have been noticed. What happened to science in *Science*?

We can now go one step further and confront Alroy's simulations with the new information about the growth of human population. The growth of the world population in the past 2,000,000 years is shown in Figure 2. It is based on my analysis (Nielsen, 2017c) of a wide range of population data (Biraben, 1980; Birdsell, 1972; Clark,1968; Cook,1960; Deevey, 1960; Durand, 1974; Gallant, 1990; Hassan, 2002; Haub, 1995; Livi-Bacci, 1997; McEvedy & Jones, 1978; Taeuber & Taeuber, 1949; Thomlinson, 1975; Trager, 1994; United Nations, 1973, 1999, 2013; US Census Bureau, 2017). The relevant part of this distribution is compared in Figure 3 with Alroy's computer simulated distribution of the growth of the hypothetical human population between 14,000 BP and 11,500 BP.

The growth rate of human population in 13,500 BP was 0.0087% (Nielsen, 2017c). My calculations are based on fitting hyperbolic distributions to the data. My growth rate is for the fitted distribution at that particular time, as displayed in Figure 2. Details of my analysis are described in my publication.

We can see that Alroy's growth rate is around 191 times larger than the growth rate for the growth of the world population at that time. Between 14,000 BP and 13,700 BP (corresponding the maximum of the simulated growth of the hypothetical human population, the world population increased by only 100,000. Alroy's population in North America increased by 1,000,000. The increase of the hypothetical population in the relatively small area in North America was 10 times larger than the increase in the growth of the world population. It is a clear aberration, which cannot be accepted.

If we look at Figure 3 we can see that while the hypothetical human population in North America was increasing fast, the growth of the world population was slow. We can also see that the overall shape of the distribution describing the growth of the hypothetical human population is manifestly and unacceptably different than the distribution describing the growth of the world population. Why should we have such a strange growth in North America? Something unusual is assumed to have been happening in North America to try to claim the human-assisted extinction of megafauna.

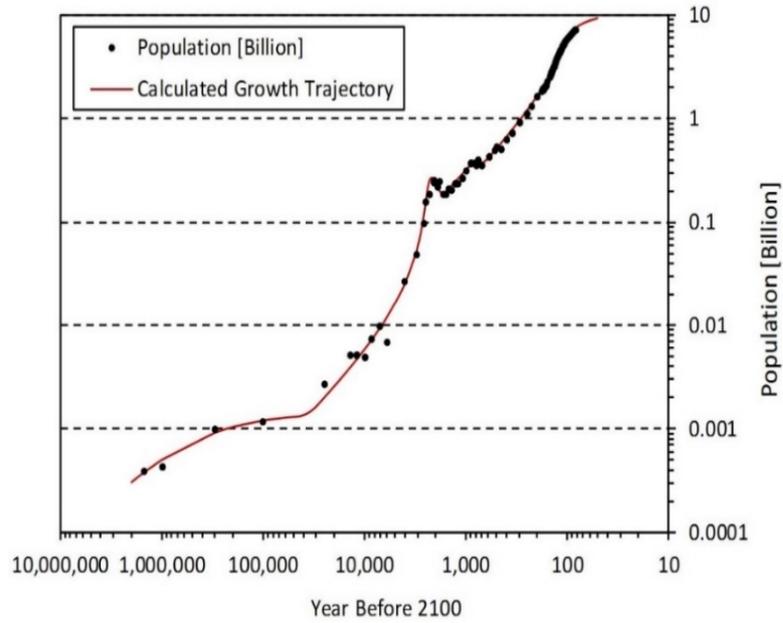

**Figure 2.** *Growth of the global human population in the past 2,000,000 years (Nielsen, 2017c) based on the analysis of a wide range of data (Biraben, 1980; Birdsell, 1972; Clark,1968; Cook,1960; Deevey, 1960; Durand, 1974; Gallant, 1990; Hassan, 2002; Haub, 1995; Livi-Bacci, 1997; McEvedy & Jones, 1978; Taeuber & Taeuber, 1949; Thomlinson, 1975; Trager, 1994; United Nations, 1973, 1999, 2013; US Census Bureau, 2017).*

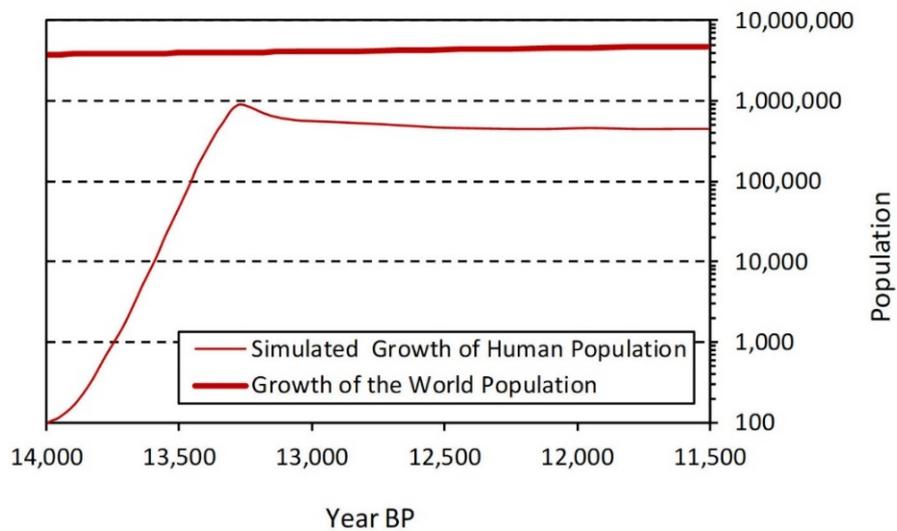

**Figure 3.** *Computer simulated distribution describing the alleged growth of human population in North America (Alroy, 2001) is compared with the growth of the world population (Nielsen, 2017c). Computer simulation is so incongruous that it has to be rejected.*

Even without the new information about the growth of the world population presented in Figures 2 and 3, Alroy should have noticed the warning signs in his own simulations. He should have noticed the unrealistically fast growth of the hypothetical human population. He should have noticed the strange overall shape of the growth trajectory. He should have noticed that the simulated growth of the hypothetical human population in North America was levelling off at a constant value, which prevailed to the present time (Alroy, 2001, p. 1895). while the real size of human population was obviously not constant. He should have noticed that there was no clear correlation between the distribution describing the growth of the hypothetical human population and the decline in the population of megafauna. He should have noticed that the two types of distributions are independent but he did not.

We can also compare his simulations with the growth of human population in Australia (Nielsen, 2017d) shown in Figure 4. Details of this analysis are described in my paper. The data are based on the records of the number of the rock shelters (Johnson, & Brook, 2011). I have combined them with the population data of Maddison (2010).

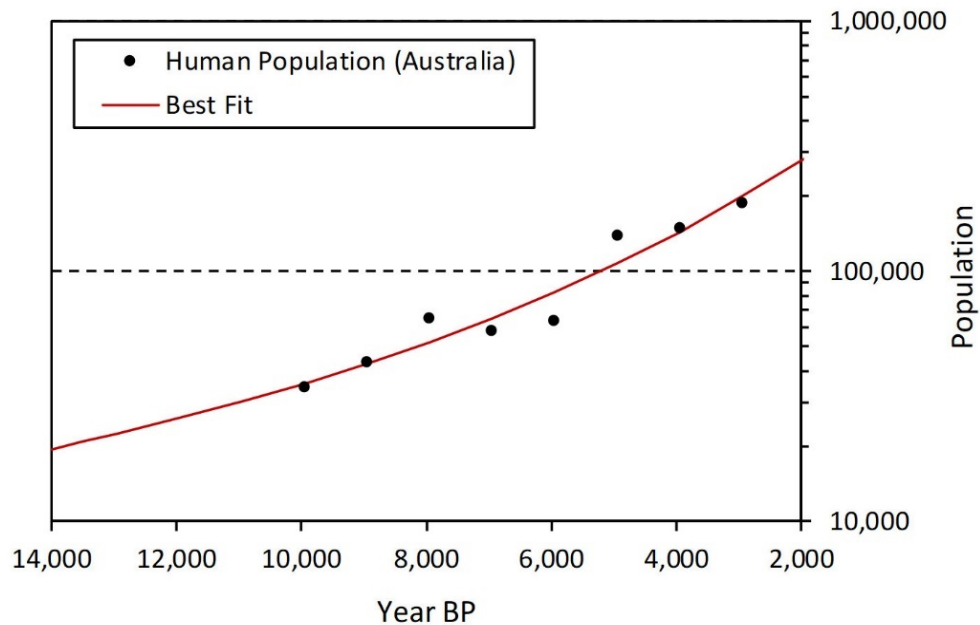

**Figure 4.** *Growth of human population in Australia (Nielsen, 2017d)*

Growth rate for the growth of population in Australia in 13,500 BP was 0.0134%. It was larger than the growth rate for the world population but only about 1.5 times larger. The growth rate for the hypothetical human population in North America was 124 times larger, clearly unacceptable. Between 14,000 BP and 13,750 BP, population in Australian increased by only around 1300. The hypothetical population in North America increased by around 1,000,000. The increase of the hypothetical human population in North America was around 770 times larger than the increase of the population in Australia during the same time.

Here also we can see that the frenetic growth of the hypothetical human population in North America is not reflected in the monotonic growth of human population in Australia. Something most unusual is assumed to have been happening in North America to try to cause the extinction of megafauna by humans but even this frenetic growth of the alleged human population cannot be used to claim that the extinction of megafauna in North America was caused by humans.

While the growth of human population in Australian was increasing steadily and monotonically, the growth of the hypothetical human population was following a strange and unusual trajectory as if designed to support the preconceived ideas of human-assisted extinction of megafauna and of reaching a carrying capacity, simulations designed not to test these ideas but to prove, in whatever way, their validity. Population data (Biraben, 1980; Birdsell, 1972; Clark,1968; Cook,1960; Deevey, 1960; Durand, 1974; Gallant, 1990; Hassan, 2002; Haub, 1995; Livi-Bacci, 1997; McEvedy & Jones, 1978; Taeuber & Taeuber, 1949; Thomlinson, 1975; Trager, 1994; United Nations, 1973, 1999, 2013; US Census Bureau, 2017) give no support to the strange distribution describing the growth of the hypothetical human population in North America generated by Alroy's computer simulations. They also give no support for the concept of reaching a constant size of population reflecting allegedly the carrying capacity. The distribution generated by Alroy is so divorced from the well-accepted and well-documented data that they belong to another world. They describe a process that never happened on our planet.

Alroy's distribution for the growth of the hypothetical human population does not simulate the growth of human population. It represents meaningless distribution that has nothing to do with the growth of human population. In his publication, he lists other examples of his computer simulations with the growth rates ranging from 1.14% to 1.92% (Alory, 2001, p.1894), all of them unacceptably large.

## Realistic simulations

Realistic simulations would have to be based on the realistic assumptions about the growth of human population. For a start, we would have to assume hyperbolic growth because trajectories describing global and regional growth of population were hyperbolic (Nielsen, 2016a, 2016b, 2017c; von Foerster, Mora & Amiot, 1960). There were some rare interruptions but the growth soon resumed its preferred hyperbolic trajectory.

We would also have to base our computer simulations on a realistic growth rate. For the hyperbolic growth, the growth rate is not constant, as for the exponential growth, but increases also hyperbolically. A reasonable assumption about the initial growth rate around 14,000 BP would be around 0.01%, preferably not greater than 0.01%. We should also assume a reasonable size of human population in North America around that year, probably not greater than a few thousand.

Under these assumptions, the growth of human population would be expected to be slow. It would probably be similar to the growth of the wold population shown in Figure 3 or for the Australian population shown in Figure 4.

We would then also have to make certain assumptions about food supply, which was probably plentiful around that time, and about the dietary habits of this early human population. It would be of course unreasonable to assume that they were eating nothing but megafauna.

Even without carrying out such computer simulations we should expect that such a small number of people, increasing so slowly, would have no impact on reducing the population of Megafauna, let alone causing the extinction of many of their species.

## Summary and conclusions

Computer simulations of the extinction of megafauna, carried out by Alroy (2001), *do not demonstrate the human-assisted extinction of megafauna*.

These simulations are fundamentally incorrect because they are based on the clearly unacceptable assumption about the growth of the hypothetical human population in North America. The assumed growth rates are about two-orders of magnitude larger than could have been reasonable expected. It is well known that the growth of population in the distant past was slow. In contrast, the assumed growth in these computer simulations was excessively fast. The produced distribution for the growth of the hypothetical human population is clearly so incongruous that it has to be rejected, which means that the simulations published by Alroy also have to be rejected as scientifically untenable.

Intentionally or unintentionally, an attempt was made to *force* the human-assisted extinction of megafauna by introducing the absurdly fast growth of the hypothetical human population. However, even this unreasonable attempt failed because it did not produce any correlation between the growth of the hypothetical human population and the distributions describing the population of megafauna. The two types of distributions are independent. For 500 years, the fast-increasing population of the hypothetical humans had absolutely no impact on the population of megafauna. But, then, for no apparent reason, the population of megafauna started to decline rapidly. However, this sudden and rapid decline is not, in the slightest way, reflected in the growth trajectory of the hypothetical human population. Based on this failed attempt we can conclude that *these computer simulations do not support the concept of human-assisted extinction of megafauna.*

The closer we analyse the concept of human-assisted extinction of megafauna, the easier it is see that this postulate is unrealistic. We have to look for far stronger forces and the obvious place to look for them is in the natural phenomena. Human impacts are now strong but they were still weak in this distant past of the Later Pleistocene. It is incorrect to extrapolate our present strong impact to the human impact in this distant past.

Computer simulations would have to be repeated by using reasonable assumptions about the growth of human population. However, it could be expected, that the growth of the hypothetical population in North America would be slow (cf. Figures 3 and 4) and that it would have no impact on the population of megafauna. Such results would be perhaps disappointing for those who believe in the human-assisted extinction of megafauna, but it would be science.

Correspondence with John Alroy is gratefully acknowledged.